\documentclass[10pt,journal,twocolumn]{IEEEtran}
\IEEEoverridecommandlockouts
\usepackage{cite}
\usepackage{amsmath,amssymb,amsfonts,booktabs}
\usepackage{algorithmic}
\usepackage{graphicx}
\usepackage{textcomp}
\usepackage{xcolor}
\usepackage{enumerate}
\usepackage{epstopdf}
\usepackage{subcaption}
\usepackage{multirow}
\usepackage{amsthm}

\usepackage{algorithm}
\usepackage{algorithmic}
\allowdisplaybreaks[4]

\begin{document}


\title{Retrieval-Augmented Generation for Mobile Edge Computing via Large Language Model}

\author{Runtao Ren,~\IEEEmembership{Student           Member,~IEEE},
        Yinyu Wu,~\IEEEmembership{Student           Member,~IEEE},
        Xuhui Zhang,~\IEEEmembership{Student Member,~IEEE},
        Jinke Ren,~\IEEEmembership{Member,~IEEE},
        Yanyan Shen,~\IEEEmembership{Member,~IEEE},
        Shuqiang Wang,~\IEEEmembership{Senior Member,~IEEE},\\
        and Kim-Fung Tsang,~\IEEEmembership{Fellow,~IEEE}

\thanks{
R. Ren is the with Shenzhen Institute of Advanced Technology (SIAT), Chinese Academy of Sciences (CAS), Guangdong 518055, China, and with the Shenzhen Future Network of Intelligence Institute (FNii-Shenzhen), the Chinese University of Hong Kong, Shenzhen, Guangdong 518172, China, and also with the Department of Information Systems, City University of Hong Kong, Hong Kong (e-mail: runtaoren@gmail.com).
(\emph{Corresponding authors: Yanyan Shen; Jinke Ren})
}

\thanks{
Y. Wu is with the SIAT, CAS, Guangdong 518055, China, and also with the University of Chinese Academy of Sciences, Beijing 100049, China (e-mail: yg.wu@siat.ac.cn).
}

\thanks{
X. Zhang and J. Ren are with the FNii-Shenzhen, the School of Science and Engineering, and the Guangdong Provincial Key Laboratory of Future Networks of Intelligence, the Chinese University of Hong Kong, Shenzhen, Guangdong 518172, China (e-mail: xu.hui.zhang@foxmail.com; jinkeren@cuhk.edu.cn).
}

\thanks{
Y. Shen, S. Wang and K.F. Tsang are with the SIAT, CAS, Guangdong 518055, China, and also with Shenzhen University of Advanced Technology, Guangdong 518055, China (e-mail: yy.shen@siat.ac.cn; sq.wang@siat.ac.cn; kftsang@ieee.org).
}

}

\maketitle

\begin{abstract}
The rapid evolution of mobile edge computing (MEC) has introduced significant challenges in optimizing resource allocation in highly dynamic wireless communication systems, in which task offloading decisions should be made in real-time. However, existing resource allocation strategies cannot well adapt to the dynamic and heterogeneous characteristics of MEC systems, since they are short of scalability, context-awareness, and interpretability. To address these issues, this paper proposes a novel retrieval-augmented generation (RAG) method to improve the performance of MEC systems. Specifically, a latency minimization problem is first proposed to jointly optimize the data offloading ratio, transmit power allocation, and computing resource allocation. Then, an LLM-enabled information-retrieval mechanism is proposed to solve the problem efficiently. Extensive experiments across multi-user, multi-task, and highly dynamic offloading scenarios show that the proposed method consistently reduces latency compared to several DL-based approaches, achieving 57\% improvement under varying user computing ability, 86\% with different servers, 30\% under distinct transmit powers, and 42\% for varying data volumes. These results show the effectiveness of LLM-driven solutions to solve the resource allocation problems in MEC systems.

\end{abstract}


\section{Introduction}

With the increasing computational demands of mobile devices, traditional cloud computing architectures face challenges such as long transmission latency and high energy consumption due to the long distances between cloud centers and mobile devices. Mobile edge computing (MEC) systems, as an emerging computation paradigm, address these issues by deploying computational and storage resources at the network edge \cite{8664595}. It significantly enhances task processing efficiency, reduces transmission latency, and provides an effective solution for latency-sensitive and computation-intensive applications, such as autonomous driving and augmented reality \cite{mec_1, 8612452}. In MEC, computation offloading is a key technology that transfers computation tasks from resource-constrained mobile devices to powerful edge servers for remote processing, thereby alleviating the computational burden on mobile devices \cite{mec_2}. 

The primary goal of computation offloading is to improve the system performance, particularly the energy consumption and end-to-end latency \cite{mec_3, 7762913}.
However, the design of offloading strategies must balance many key factors, including network bandwidth, task priority, offloading ratio, and the load conditions of edge servers \cite{mec_6, Zhang2023Learning}.
In particular, these variables are coupled with each other, making it challenging to achieve optimal resource allocation and meet the diverse requirements of intelligent applications, such as image processing and online video analysis.

In scenarios where multiple mobile devices share limited communication and computational resources, ensuring fair and efficient resource allocation is essential. Previous works have proposed many intelligent frameworks to achieve this goal \cite{mec_8, 8714026}.
A notable approach is the utilize of deep learning (DL) algorithms for dynamic resource allocation.
The application of DL algorithms in MEC systems offers several advantages. First, DL methods can analyze historical data to predict user demand patterns, enabling proactive resource allocation. Second, by efficiently integrating computational and communication resource management, DL methods can reduce system latency and improve resource utilization across diverse and dynamic network environments 
\cite{irsrelate, 9214878, 10174680, 10293163}.

Although DL methods can enhance system performance to a certain extent, they still suffer from three limitations \cite{8660505, peng2024survey}: (1) DL methods are difficult to generalize to scenarios that have not been encountered; (2) DL methods are usually trained with fixed objectives and predefined architectures, and are difficult to adapt to highly dynamic MEC systems; (3) DL methods are usually black boxes. It is difficult to track or verify the logic behind offloading decisions. Due to these issues, traditional DL methods show limitations in addressing the dynamic and complicated requirements of MEC systems. These limitations stem from their reliance on fixed architectures, predefined objectives, the lack of interpretability, and their inability to incorporate external knowledge or adapt to real-time changes effectively. With the growing complexity of MEC systems, there is an urgent need for a more adaptive, user-centered, and context-aware approach to computation offloading.

Recent advancements in artificial intelligence (AI), particularly large language models (LLMs), offer a promising solution to these challenges \cite{jiang2024large}. LLMs can bridge the gap between user intents and system-level decisions by understanding natural language inputs. While LLMs excel in reasoning and natural language understanding, they lack the capability to retrieve and incorporate real-time information from external sources, such as network bandwidth, edge server loads, and user priorities \cite{he2024large}. This limits their effectiveness in handling highly dynamic and resource-constrained environments that necessitate real-time decision-making. To further improve the adaptability and reliability of LLM, it is necessary to incorporate new techniques in complex MEC systems \cite{intro11}.

In this paper, we employ retrieval-augmented (RA) techniques to minimize the overall latency of the MEC system, which combine the reasoning capability of LLMs with the contextual adaptability of retrieval mechanisms to form the state-of-the-art retrieval-augmented generation (RAG) method. RAG allows LLMs to dynamically retrieve relevant and up-to-date information during decision-making, bridging the gap between general reasoning and real-time adaptability \cite{lewis2020retrieval}. This integration addresses two fundamental challenges: (1) dynamic decision-making: By providing LLMs with accurate and real-time information about the system (e.g., network bandwidth, edge server loads), RAG ensures that the output offloading decisions are contextually relevant and aligned with the current network conditions and user demands; (2) enhanced interpretability: RAG enables the creation of traceable decision paths, where the sources of retrieved information can be validated. In this paper, we explore the combination of RAG and MEC optimization. The main contributions are summarized as follows.
\begin{itemize}
\item We for the first time apply RAG to MEC systems. Specifically, we formulate a latency minimization problem to optimize the task execution efficiency in an MEC system. Then, we introduce a novel RAG-based framework to tackle the challenges of dynamic resource allocation and real-time decision-making, enabling adaptive and efficient offloading strategies.

\item By integrating real-time retrieval mechanisms, our framework enables LLMs to generate adaptive and user-specific offloading strategies that efficiently respond to dynamic network and computing conditions, thereby improving the overall performance and resource utilization efficiency of the MEC system.

\item Through extensive experiments, we demonstrate that the proposed RAG framework outperforms traditional DL-based methods, achieving 57\% improvement under varying user computing ability, 86\% with different servers, 30\% under distinct transmit powers, and 42\% for varying data volumes in optimizing latency.
\end{itemize}

The remainder of this article is organized as follows. Section II discusses the related works. Section III introduces the system model, 
including the communication and computation models. In Section IV, we formulate the latency minimization problem and introduce the solution via LLM inference. Section V evaluates the performance of the proposed solution through experiments. Finally,
Section VI concludes this article.

\section{Related Works}

\subsection{Classical Computation Offloading}

Computation offloading refers to a technique that involves transferring computation tasks from resource-constrained mobile devices to powerful edge server to alleviate the computation burden on the devices. The primary goal of computation offloading is to optimize resource allocation and task scheduling to improve the system performance, such as latency and energy consumption.
Current research on computation offloading primarily addresses resource constraints in dynamic MEC systems through optimization methods. For instance, Zamzam et al. \cite{relate_mec_2} adopted a game-theoretic approach to resolve resource competition in a multi-user scenario, demonstrating the ability to allocate limited resources efficiently.
Huang et al. \cite{relate_mec_3} developed a joint optimization framework that considers both network congestion and latency, reducing energy consumption and task execution times by simultaneously optimizing data transmission paths and offloading strategies. Zhao et al. \cite{relate_mec_4} introduced an adaptive mechanism that dynamically adjusts offloading decisions to meet diverse task demands, such as latency requirements and bandwidth availability.

Despite progress, current solutions often rely on traditional rule-based or model-based optimization approaches, which lack the adaptability required for dynamic MEC systems. Moreover, they fail to incorporate real-time contextual communication information, such as current user priorities or the operational state of the network and edge servers, leading to decisions that do not align with the system’s current needs. These underline the limitations of existing computation offloading methods, which often prioritize simplicity over adaptability and fail to address the dynamic, heterogeneous, and large-scale nature of MEC systems. Addressing these challenges requires a more intelligent and context-aware approach, capable of dynamically adjusting to environmental changes while ensuring efficient allocation. 

\subsection{Reinforcement Learning for Computation Offloading}
With the rapid development of reinforcement learning (RL), many studies have applied RL to computation offloading problems, leveraging its adaptability and dynamic decision-making capabilities to address complex offloading tasks. In the optimization of computation offloading, RL-based approaches optimize task allocation and resource management by continuously adapting to changing network and system conditions. For instance, a model-free RL-based offloading algorithm was proposed in \cite{8444467}, which helps mobile users learn their long-term task offloading strategies to maximize their long-term utilities.
Yao et al. \cite{8657779} utilized a multi-agent RL algorithm to find an optimal resource management and pricing strategy for the computation offloading.
Alfakih et al. \cite{9039672} studied a multi-server offloading system and proposed an RL-based algorithm to minimize the system cost function by making the optimal offloading decision.
Moreover, Wu et al. \cite{9117034} proposed a collaborative RL-based routing algorithm for a multi-access vehicular edge computing system with a low communication overhead.

Despite the strong potential of RL in computation offloading, its training process faces challenges such as high computational resource demands and difficulties in data collection, which limit its application.

\subsection{Deep Learning for Computation Offloading}
DL leverages deep neural networks (DNNs) to model complex data distributions. In computation offloading, DL methods are often employed to enhance task allocation and resource management, leveraging their ability to learn patterns from large datasets.
For instance,
Huang et al. \cite{dl_mec_1} proposed a deep RL (DRL)-based strategy to optimize the system performance, which dynamically adjusts offloading policies based on network conditions and the computational capacities of mobile devices. In addition, Sadiki et al.
\cite{relate_mec_5} proposed a DRL-based computation offloading strategy, training an RL agent to dynamically adjust offloading decisions while balancing latency and energy consumption. In this approach, the agent interacted with the environment and optimized offloading decisions based on reward signals.
Similarly, Zhang et al. utilized a hybrid deep Q-network (DQN) and the block coordinate descent method to optimize data offloading decision and resource allocation, significantly enhancing system utility \cite{cyberzhang23}.
Moreover, \cite{relate_mec_drl_1} introduced a joint optimization framework for computation offloading and resource allocation in a dynamic multiuser MEC system. The proposed approach focused on minimizing energy consumption while considering latency constraints and heterogeneous task requirements.
On the other hand, Wu et al. \cite{10734319} focused on integrating MEC with AI generation content services, aiming to minimize the latency of each service to enhance the quality of service for mobile users.

While DL-based methods demonstrate great potential, they also exhibit significant shortcomings. For example, DL methods are typically trained on static datasets, limiting their ability to adapt to real-time changes in dynamic network conditions and user demands. Furthermore, DL-based methods often operate as black boxes, making their decision processes difficult to interpret, which is problematic in applications requiring trust and transparency.
These limitations underscore the challenges of using DL methods in computation offloading.

\subsection{Genrative Learning for Computation Offloading}
The limitations of RL and DL based methods underscore the need for innovative approaches that provide greater adaptability, interpretability, and context-awareness for computation offloading. Generative AI, a paradigm distinct from RL and DL, offers a promising solution. Generative AI models are designed to produce outputs, such as texts and images, based on contextual understanding and reasoning. This enables them to process ambiguous inputs, infer intent, and generate adaptive outputs aligned with dynamic conditions \cite{fui2023generative}.

LLMs are powerful generative AI models that can process and generate human-like text, solve complex problems, and adapt to a wide range of applications. However, LLMs lack direct access to up-to-date system information, such as real-time network conditions, edge server loads, and user-specific constraints, which are critical for making accurate and context-aware decisions in dynamic MEC systems. Therefore, they fail to align with the immediate needs of the system. To overcome these gaps, RAG integrates LLMs with a retrieval mechanism, enabling them to access and leverage real-time information during the decision-making process.  In this study, we propose an innovative RAG-enhanced MEC optimization framework, which, for the first time, applies RAG techniques to the computation offloading problem. By integrating the reasoning power of LLMs with real-time retrieval mechanisms, this framework addresses the critical challenges of adaptability, context-awareness, and interpretability in MEC systems.

\section{System Model and Problem Formulation}

\begin{figure}[htbp]
	\centering
	\includegraphics[width=3.5in]{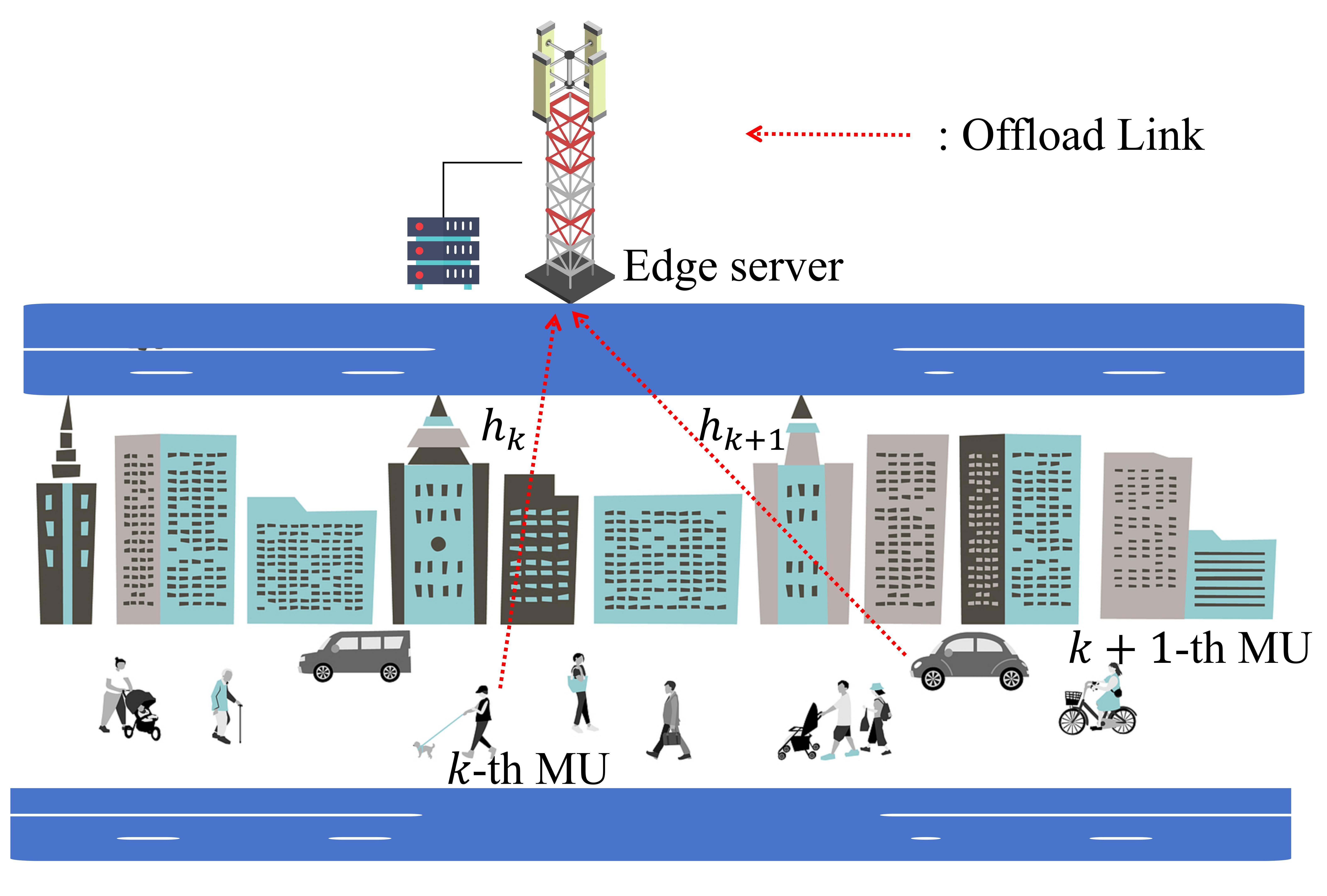}
	\caption{Multiuser MEC system model.}
	\label{fig:system}
\end{figure}

As shown in Fig. \ref{fig:system}, we consider a MEC system with one edge server and $K$ mobile users (MUs).
The set of the MUs and the time slots are given by $\mathcal{K} = \{1,\ldots,K\}$, and $\mathcal{T} = \{ 1, \ldots, T\}$, respectively.
Each MU has a computation task at the beginning of each time slot, which can be divided into two parts, with one part being processed locally and the other offloaded to the edge server for processing.
During the computation services, a time block is divided into $T$ time slots, where each time slot has a duration of $\tau$ seconds.
The data volume of the computation task of the $k$th MU at time slot $t$ is denoted as $D_k (t)$.

\subsection{Local Computing Model}
Let $\alpha_k (t) \in [0,1]$ denote the task offloading ratio, where the part of $\alpha_k (t)D_k (t)$ is offloaded to the edge server for remote processing, and the other part $(1-\alpha_k (t))D_k (t)$ is processed locally.
Let $f^{\rm{}}_{m}$ denote the computing cycles per second of the $k$th MU, $\phi$ denote the computing cycles required to process one-bit data. Then, the latency for processing the task at the $k$th MU is given by
\begin{equation}
L_{k}^{\mathsf{local}}(t) = \frac{ (1-\alpha_k^{\rm{}}(t)) \phi D_{k}(t)}{f^{\rm{}}_{k}},
\label{latencylocal}
\end{equation}
Accordingly, the energy consumption of the $k$th MU can be expressed as
\begin{equation}
E_{k}^{\rm{}}(t) = \kappa_{\rm{comp}} (f^{\rm{}}_{k})^{2} (1-\alpha_k^{\rm{}}(t)) \phi D_{k} (t),   
\end{equation}
where $\kappa_{\rm{comp}}(f^{\rm{}}_{k})^{2}$ denotes the energy consumption per computing cycle, and $\kappa_{\rm{comp}}$ is the energy coefficient \cite{7524497, 8849964}.

\subsection{Channel Model}
We assume that all MUs have a time quasi-varying locations. Specifically, each MU keeps fixed within one time slot, and has different positions across time slots. The trajectory of all MUs can be given by a certain map or be designed according to the road environment \cite{9547273, 10683327}.
Specifically, the position of the $k$th MU at time slot $t$ is given by $[ q_k^x(t),  q_k^y(t) ]$, where $q_k^x(t)$ and $q_k^y(t)$ are the two-dimensional (2D) coordinates.
Therefore, the distance between the $k$th MU and the edge server is given by
\begin{equation}
d_{k} (t) = \sqrt{
\left( q_k^x(t) -  q_{\rm{MEC}}^x \right)^2
+ \left( q_k^y(t) -  q_{\rm{MEC}}^y \right)^2
+ h_{\rm{MEC}}^2},
\end{equation}
where $q_{\rm{MEC}}^x$ and $q_{\rm{MEC}}^y$ are the fixed 2D coordinates of the edge server, and $h_{\rm{MEC}}$ denotes the fixed height of the edge server.

In this paper, we consider a channel model with both large-scale and small-scale fading. The channel gain between the $k$th MU and the edge server can be expressed as
\begin{equation}
     h_{k} (t) = g_{k} (t) \tilde{h}_{k} (t),
\end{equation}
where $g_{k}(t)$ and $\tilde{h}_{k}(t)$ represent the large-scale fading coefficient and small-scale fading coefficient from the $k$th MU to the edge server. Accordingly, $g_{k}(t)$ and $\tilde{h}_{k}(t)$ can be expressed as
\begin{equation}
    g_{k} (t) = \frac{g_0}{d^2_{k}(t)},
\end{equation}
\begin{equation}
    \tilde{h}_{k} (t) = \left( \sqrt{ \frac{\kappa}{\kappa+1} } + \sqrt{ \frac{1}{\kappa+1} }\Bar{h}_k (t)  \right)^2,
\end{equation}
where $g_0$ denotes the channel gain at the reference distance (i.e., 1 $\rm{m}$), $\Bar{h}_k (t)$ follows complex Gaussian distribution with zero mean and unit variance, i.e.,
$\Bar{h}_k (t) \sim \mathcal{CN}(0, 1)$, and $\kappa$ is the Rician factor \cite{9453748, 10606316}.

\subsection{Offloading Model}
Similar to \cite{9133107, 9446552, 9673750}, we consider a general data offloading model, where all MUs share the same bandwidth resource. In this model, the offloading signals transmitted by other MUs will be regarded as interference for a specific MU.
According to the channel model, the data offloading rate from the $k$th MU to the edge server can be expressed as
\begin{equation}
    r_k (t) = B \log_2 \left ( 1 + \frac{p_k(t) h_k (t)}{\sum_{l \in \mathcal{K},l \neq k}p_l(t) h_l (t) + \sigma^2 }
    \right ),
\end{equation}
where $B$ denotes the system bandwidth, $p_k(t)$ represents the transmit power of the $k$th MU, and $\sigma^2$ is the noise power. 
Accordingly, the data offloading latency can be expressed as
\begin{equation}
    L_k^{\mathsf{off}} (t) = \frac{\alpha_k(t)D_k(t)}{r_k (t)}.
    \label{latencyoff}
\end{equation}
Meanwhile, the energy consumption of the $k$th MU for data offloading is given by
\begin{equation}
    E_k^{\mathsf{off}} (t) = 
    p_k(t) L_k^{\mathsf{off}} (t).
\end{equation}

We assume that the total computational capability of the edge server is $F$, which is assigned for processing the task offloaded by all MUs.
Let $\beta_k (t) \in [0,1]$ denote the computational capability ratio allocated to the $k$th MU. Therefore, $\sum_{k=1}^K \beta_k (t) \leq 1$. As a result, the computation latency for the edge server to process the task from the $k$th MU can be expressed as
\begin{equation}
    L_k^{\mathsf{MEC}}(t) = \frac{\alpha_k(t)\phi D_{k}(t)}{\beta_k (t)F}.
    \label{latencymec}
\end{equation}

\subsection{Problem Formulation}
According to Eqs. \eqref{latencylocal}, \eqref{latencyoff}, and \eqref{latencymec}, the overall latency of the $k$th MU is given by
\begin{equation}
\label{time_latency}
    L_k (t) = \max \{L_{k}^{\mathsf{local}}(t), L_{k}^{\mathsf{off}}(t) + L_{k}^{\mathsf{MEC}}(t)\}.
\end{equation}
In this work, we aim to minimize the average latency of all MUs, by jointly optimizing the task offloading ratio, $\boldsymbol{\alpha} = \{\alpha_k(t)\}$, the computational capability ratio $\boldsymbol{\beta} = \{\beta_k(t)\}$, and the transmit power allocation of all MUs $\boldsymbol{p} = \{p_k(t)\}, \forall k, \forall t$. Therefore, the latency minimization problem can be formulated as
\begin{subequations}\label{P1}
\begin{flalign}
\label{question}
  \textbf{P1}\ &\min_{\boldsymbol{\alpha}, \boldsymbol\beta ,\boldsymbol{p}} \ \ \frac{1}{T}\frac{1}{K}  \sum_{t\in \mathcal{T}}  \sum_{k\in \mathcal{K}} L_{k} \left ( t \right ) \tag{\ref{P1}} \\
 {\rm{s.t.}}  \quad & 0\le p_k (t) \le P_{k}^{\max}, \quad \forall k\in \mathcal{K},   \label{P1a}\\
& \alpha _{k} \left ( t \right ) \in \left [ 0,1 \right ],  \quad \forall k\in \mathcal{K},\ \forall t\in \mathcal{T},\label{P1b}\\
& \beta _{k} \left ( t \right ) \in \left [ 0,1 \right ],  \quad \forall k\in \mathcal{K},\ \forall t\in \mathcal{T},\label{P1c}\\
& \sum_{k=1}^K \beta_k \left ( t \right ) \leq 1,\ \forall t\in \mathcal{T}, \label{P1d}\\
& \sum_{t=1}^T \left (E_k^{\mathsf{}} (t) + E_k^{\mathsf{off}} (t) \right )
\leq E_k^{\max},\ \forall k\in \mathcal{K},\label{P1e}
\end{flalign}
\end{subequations}
where $E_k^{\max}$ is the maximum energy capacity of the $k$th MU,
\eqref{P1a} represents the transmit power limitation of all MUs, \eqref{P1b} denotes the range of the offloading ratio of all MUs, \eqref{P1c} is the range of allocated computational capability ratio to each MUs by the edge server, \eqref{P1d} represents that the total allocated computational capability of the edge server cannot exceed its maximum computational capability, and \eqref{P1e} represents that the total energy consumption of each MU cannot exceed its maximum energy capacity.

In the objective function \eqref{P1}, it can be observed that the transmit power is coupled with the data offloading rates of all MUs.
Hence, problem \textbf{P1} is a non-convex problem.
Meanwhile, the MUs moves during the time block, and the channel conditions are continuously time-varying.
As a result, problem \textbf{P1} is hard to be solved by conventional optimization algorithms.
Many recent works have utilized learning based methods (e.g., DL and RL) to solve such problems.
However, these methods are difficult to generalize to highly dynamic MEC systems.
Meanwhile, their decisions lack interpretability. Differently, we apply the RAG framework to solve problem \textbf{P1}. From a mathematical perspective, RAG enhances both the optimization process and the decision-making interpretability in the following ways: (1) Dynamically retrieving variables based on real-time conditions to handle time-varying conditions. (2) Reducing the complexity of coupled variables by narrowing the solution space with retrieved priors. In addition, from an engineering perspective, the RAG enhances decision-making interpretability by providing traceable paths between inputs and outputs. Furthermore, the RAG method adapts quickly to real-time scenarios without retraining. Through these mechanisms, RAG offers a powerful and practical solution for optimizing MEC systems in dynamic and complex environments.

\section{Solution via LLM}
\subsection{Preliminaries of RAG}
RAG is a hybrid architecture that combines information retrieval techniques with the generative capabilities of LLMs \cite{IR3}. RAG operates through a two-stage process, as shown in Fig.2. The first stage uses the retrieval module to obtain relevant information from the knowledge base, and the second stage uses the generation module to generate the final answer based on these retrieval results.

\begin{figure}[htbp]
	\centering
	\includegraphics[width=3.5in]{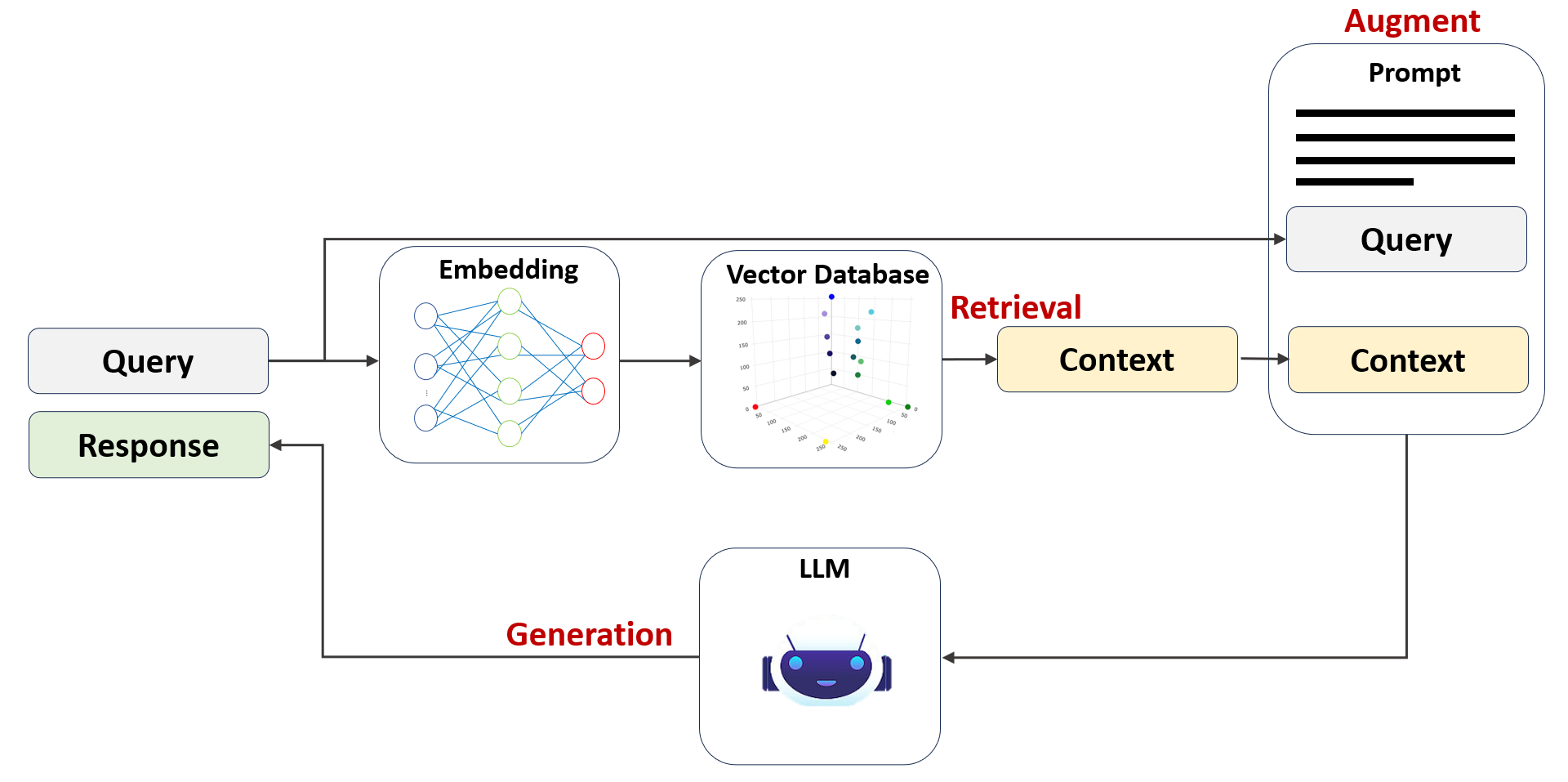}
	\caption{The framework of RAG.}
	\label{fig:Generate_prompt}
\end{figure}

The retrieval component of RAG relies on dense retrieval techniques, which use vector representations to capture the semantic relationship between user queries and pre-stored configurations in a knowledge base. Unlike traditional keyword-based methods, dense retrieval encodes both queries and documents into high-dimensional vectors, enabling more accurate and context-aware matching. The Bi-encoder model is one of the core architectures for implementing dense retrieval \cite{xiao2024c}, which is usually composed of two independent encoders: one for processing queries and the other for processing documents. Both encoders use DNN (e.g., Transformers) to convert queries and documents into vector representations \cite{IR2}. In this way, the Bi-encoder model is able to calculate the similarity between queries and documents in a shared vector space, thereby retrieving the most relevant configurations. The Bi-encoder model usually consists of the following modules (as shown in Fig.3):

\begin{itemize}
\item Query encoder: The encoder receives the input query and converts it into a high-dimensional vector representation. 

\item Document encoder: The document encoder encodes a predefined knowledge base and converts them into a vector representation. 
\end{itemize}
By calculating the similarity between the query vector and the document vector (usually using cosine similarity), the Bi-encoder model can retrieve the documents most relevant to the query.
\begin{figure}[htbp]
	\centering
	\includegraphics[width=3in]{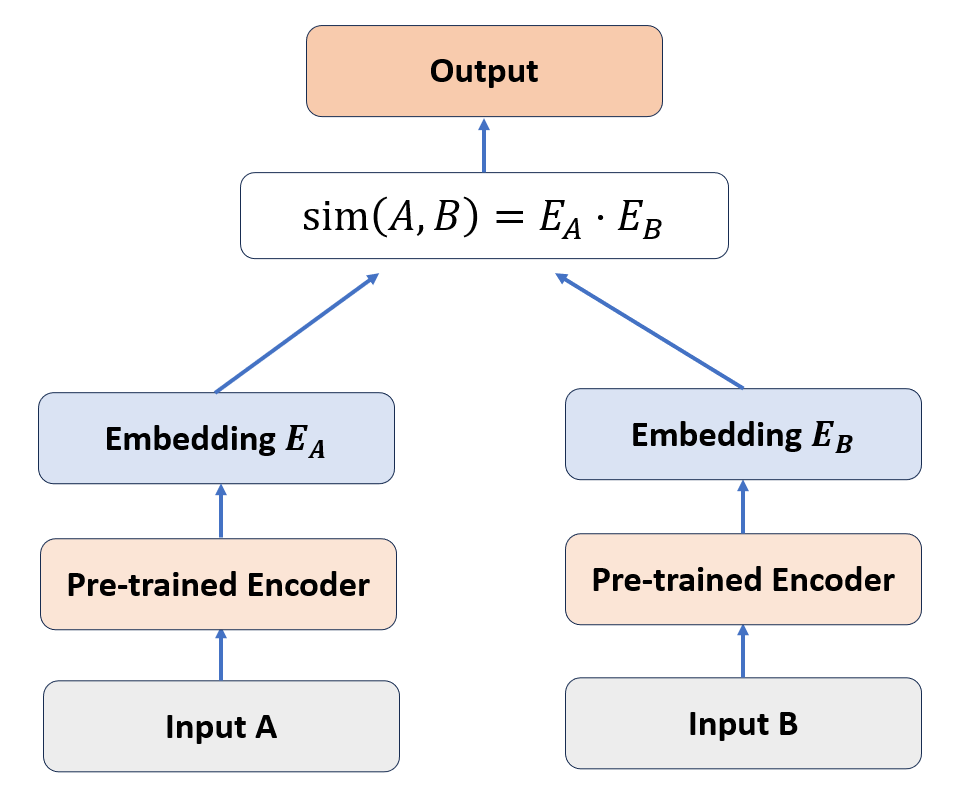}
	\caption{The framework of Bi-encoder.}
	\label{fig:Generate_prompt}
\end{figure}

The second stage of RAG involves leveraging the generative capabilities of LLMs. In this architecture, LLMs act as the reasoning and decision-making engine, synthesizing retrieved information with the task-specific query to produce optimized solutions. LLMs in RAG are not confined to predefined rules or static algorithms. Instead, they dynamically adapt to the input context by integrating retrieved knowledge with their pre-trained reasoning capabilities. By integrating retrieved information, LLMs enhance their interpretability and decision-making efficiency, producing more accurate and practical optimization outputs for MEC systems.

In the MEC system, RAG can be applied to problems such as dynamic task offloading and resource allocation. Specifically, the first stage of the RAG architecture can retrieve historical data or best practices related to specific network states (such as communication latency, bandwidth limitations, computational capability, etc.). Then in the second stage, the LLM generates corresponding resource allocation strategies or task scheduling decisions based on these data.

By combining RAG with Bi-encoder and LLMs, MEC systems can not only utilize external knowledge in the generation process, but also obtain accurate and timely optimization suggestions through efficient dense retrieval technologies, thereby improving overall system performance and decision-making efficiency.

\subsection{Solution}
This study adopts the RAG method to optimize the data allocation ratio, transmit power allocation, and computational capability allocation in the communication decision-making process, to minimize total latency. The RAG framework stores user configuration information of computing capability in a vector knowledge base, retrieves relevant configurations of computing capability based on the identity of MUs during data transmission, and ultimately generates optimization decisions using an LLM. The entire process includes three modules: storage, retrieval, and generation, which work together to solve the problem \textbf{P1}, as shown in Fig. \ref{fig:RAG_process}. The function of each module can be described as follows:

\begin{figure*}[htbp]
	\centering
	\includegraphics[width=6.5in]{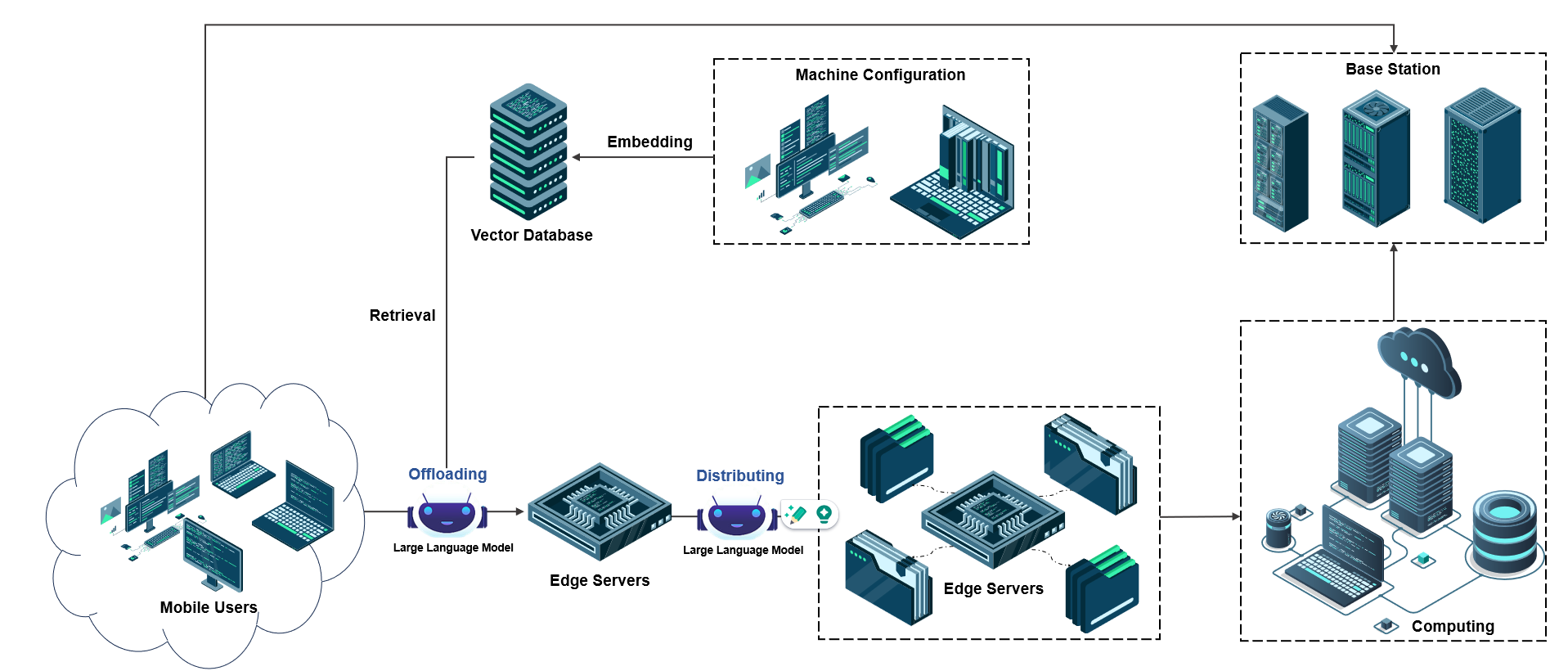}
	\caption{The framework of RAG for the MEC system.}
	\label{fig:RAG_process}
\end{figure*}

\textbf{Storage Module}: The storage module is responsible for storing all MUs's configuration information of computing capability in the vector database. Firstly, it converts each user's computing cycles per second $f_{m}$ into a vector using the embedding model:
\begin{equation}
v_{k} = \emph{Encode}(f_{k}).
\end{equation}
Then all MUs' vector representations of computing capability $\textbf{V}={(v_{1}, v_{2}, ..., v_{n})}$ are stored in an efficient vector database to support fast similarity retrieval.

\textbf{Retrieval Module}: The retrieval module retrieves relevant computing capability information based on the identity information of the MUs. Initially, the system generates a query \emph{q} based on the identities $K={(k_{1}, k_{2}, ..., k_{n})}$ and data volume $D={(D_{1}, D_{2}, ..., D_{n})}$ of all MUs. Then the Bi-encoder model is utilized to retrieve the information of local computational capability \emph{C} related to the sending MUs:
\begin{equation}
    C = \text{Sim}(q, v_k) = \frac{q \cdot v_k}{\|q\| \|v_k\|}.
\end{equation}

\textbf{Generation Module}: The generation module utilizes the retrieved data $(K, D, C)$ and current communication status $(B, \sigma^2)$ to generate optimized decisions through the LLM. The module first combines the retrieved configuration information of computing capability \emph{C}, the computing capability of edge server \emph{F}, the amount of data sent $(K, D)$, and the communication parameters $(B, \sigma^2)$ to merge into the prompt $P=(K||D||C||F||B||\sigma^2)$, as shown in Fig. \ref{fig:Generate_prompt}. Then the system inputs the prompt $P$ into the LLM to output optimization decisions:
\begin{equation}
L_k = \emph{LLM}(P).
\end{equation}

\begin{figure}[htbp]
	\centering
	\includegraphics[width=3.5in]{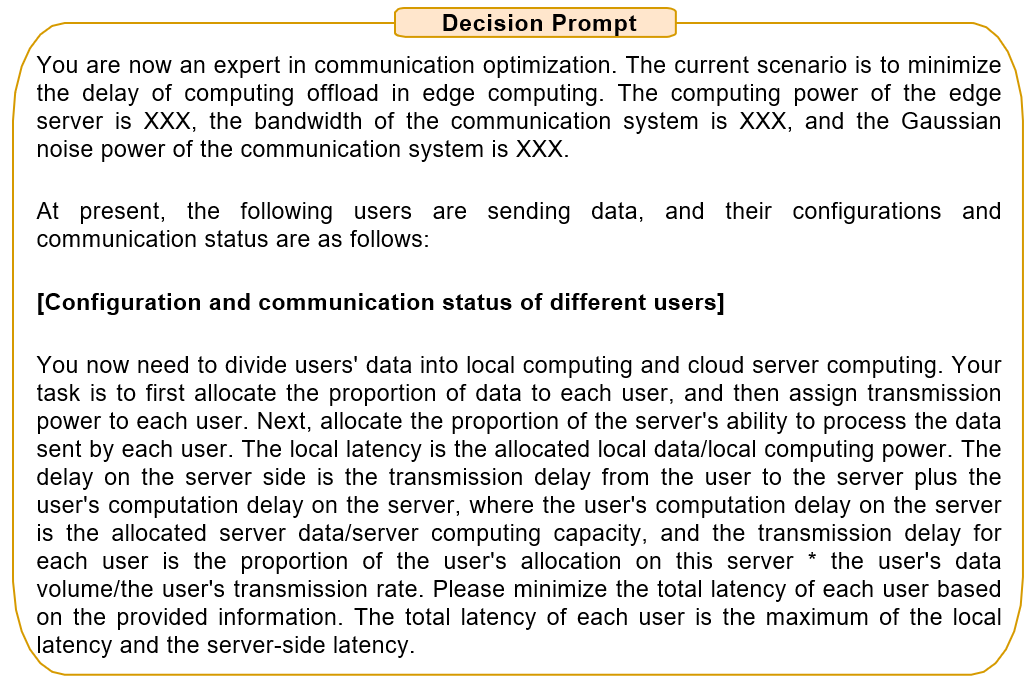}
	\caption{The prompt of LLM for decision}
	\label{fig:Generate_prompt}
\end{figure}

\section{Experiments}
\subsection{Experimental Setup}
We consider a MEC system with $10$ users and one edge server, which are randomly distributed in a square area of $300\times 300 \text{m}^2$. 
The computational capability of MUs is randomly generated within the range $[0.5, 2]$ \text{GHz} and remains fixed throughout the simulation.  The task data size is set within the range $[0.5, 5]$ \text{Mbits}, and the processing cycles per bit are randomly initialized within $[500, 1500]$ \text{cycles/bit}.
The computational capability of the edge server is $30 \text{GHz}$ and requires $900 \text{cycles}$ to process one-bit data. The communication channel adopts a free-space path loss model with a reference gain of $g_0 = 10^{-5}$, combined with Rician fading characterized by a Rician factor of $\kappa = 50$. The noise power is set to $10^{-10} \text{W}$, and the communication bandwidth is $10 \text{MHz}$. The maximum allowable transmit power for each MU is $P_{\text{max}} = 2 \text{W}$.

\subsection{Datasets and Metrics}
To evaluate the performance of the proposed RAG-based optimization method under various configurations, we choose the following datasets for testing: (1) different server computational capabilities (DSCC), where server computational capacities are varied to investigate the impact of edge server capabilities on system performance; (2) different user data volumes (DUSD), where the size of tasks is modified to examine system adaptability; (3) different user power (DUP), where transmit power is altered to assess energy constraints' influence on system performance; (4) different user computational capabilities (DUCC), where the maximum computational capacities of MUs are varied to understand their role in task offloading processes.
Each dataset contains 10 time slots, with parameter ranges divided to benchmark system performance, as shown in Table \ref{tab:parameter_table}. For measuring the performance of the proposed method, we use three metrics:

\begin{table*}[ht]
\renewcommand{\arraystretch}{1.5} 
\centering
\caption{Dataset Parameters and Their Descriptions}
\label{tab:parameter_table}
\begin{tabular}{|c|c|c|}
\hline
\textbf{Dataset} & \textbf{Parameter Description} & \textbf{Parameter Size} \\
\hline
\text{DSCC} & \text{Server computing capability ($\frac{\phi}{F}$ s/bit)} & [1e$^{-8}$, 2e$^{-8}$, 3e$^{-8}$, 4e$^{-8}$, 5e$^{-8}$] \\
\hline
\text{DUSD} & \text{User data volume (Mbit)} & [0.5, 1], [1.5, 2], [2.5, 3], [3.5, 4], [4.5, 5] \\
\hline
\text{DUP} & \text{User transmit power (W)} & [0.75, 1], [1, 1.25], [1.25, 1.5], [1.5, 1.75], [1.75, 2] \\
\hline
\text{DUCC} & \text{User computing capability ($\frac{\phi}{f_k}$ s/bit)} & [0.5e$^{-6}$, 1e$^{-6}$], [1e$^{-6}$, 1.5e$^{-6}$], [1.5e$^{-6}$, 2e$^{-6}$], [2e$^{-6}$, 2.5e$^{-6}$], [2.5e$^{-6}$, 3e$^{-6}$] \\
\hline
\end{tabular}
\end{table*}

(1) \textbf{Mean reciprocal rank} (MRR). For each query, MRR evaluates the system accuracy by viewing the highest ranked relevant information \cite{MRR}. Specifically, it is the average of the reciprocal of these rankings in all queries. In edge computing optimization, we hope that the system can efficiently find the best configuration. Therefore, MRR is the core indicator to evaluate the sorting effect. It can be calculated using the following formula: 
\begin{equation}
\text{MRR} = \frac{1}{Q} \sum_{i=1}^{Q} \frac{1}{\text{rank}_i},
\end{equation}
\noindent
where $Q$ is the number of queries and $\text{rank}_i$ is the rank position of the first relevant data for the $i$-th query.

(2) \textbf{Hit rate} (HR). HR evaluates the model's ability to find the correct answer among the Top-k results \cite{MRR}. This indicator measures whether the retrieval method can find the corresponding target information in the database. Its calculation formula is:
\begin{equation}
\text{HR} = \frac{\text{Number of hits in top-k}}{\text{Total queries}}.
\end{equation}

(3) \textbf{Latency}. Latency is a direct indicator to measure the system performance, which reflects the optimization effect of the system under different experimental conditions. This indicator is determined by the local computing latency, local and edge task offloading latency, and edge computing latency, as shown in Eq. (\ref{time_latency}). 


\subsection{Baselines}
We compare the following baseline methods for performance comparison.

\textbf{DQN method}: This approach uses DQN based on the RL to dynamically make decisions on task offloading strategies. The DQN learns environmental attributes and generates optimal task offloading strategies through a large amount of training time and data. The state space includes user task data volume, channel gain, user computing capability, and server computing capability. The action space comprises user offloading ratio, transmit power, and server computing capability, each discretized into five levels, yielding $50$ action combinations. The DQN's Q-network features an input layer and two hidden layers.

\textbf{Deep deterministic policy gradient (DDPG) method}: This approach utilizes an actor-critic architecture within RL to optimize task offloading strategies in continuous action spaces. The state space includes task data volume, channel gain, MUs' computing capability, and server computing capability, while the action space comprises continuous values for the data offloading ratio, transmit power, and server resource allocation, constrained to $[0, 1]$. The actor network consists of an input layer followed by three hidden layers with dimensions $128 \rightarrow 256 \rightarrow 128$ using ReLU activation, and an output layer with a sigmoid activation to ensure the action values remain in the range $[0, 1]$.

\textbf{Proximal policy optimization (PPO) method}: This approach adopts a stochastic policy approach to optimize offloading strategies. The state space remains the same, while the continuous action space outputs actions sampled from a Gaussian distribution, parameterized by the actor network, which has three hidden layers $128 \rightarrow 256 \rightarrow 128$ with ReLU activation. It outputs both the action mean (mapped to $[-1, 1]$ using a sigmoid function) and the log standard deviation, clamped within $[-20, 2]$. The critic network in PPO estimates the state value $V(s)$ using an input layer, three hidden layers $128 \rightarrow 256 \rightarrow 128$ with ReLU activation, and an output layer that produces a scalar value.

\textbf{RAG method}: This approach utilizes the RAG framework of LLMs to address edge computing optimization problems. First, context information is retrieved from relevant data. Then, a LLM is used to decide task offloading and calculation allocation without massive training. The experiments of this method are done by the llamaindex \footnote{https://www.llamaindex.ai/}. For the retrieval process, we set the chunk size=512 tokens and Top-k=5. And we test three embedding models in the retrieval module: bge-large-en-v1.5\footnote{https://huggingface.co/BAAI/bge-large-en-v1.5}, bge-base-en-v1.5\footnote{https://huggingface.co/BAAI/bge-base-en-v1.5}, and bge-small-en-v1.5\footnote{https://huggingface.co/BAAI/bge-small-en-v1.5}. For the generation module, we utilize GPT-4o\footnote{https://platform.openai.com/docs/models/gp} and Qwen2.5-turbo\footnote{https://qwen2.org/qwen2-5-turbo/} to synthesize task offloading strategies.

\subsection{Evaluation}
We evaluate the performance of the proposed RAG-based method using three embedding models: bge-large-en-v1.5, bge-base-en-v1.5, and bge-small-en-v1.5, across four datasets (DSCC, DUSD, DUP, DUCC). Table \ref{tab:embedding_table} shows the retrieval result of different models. All models show good performance, with hit rates ranging from 0.855 to 0.948 and MRR values between 0.846 and 0.948.

\begin{table}[ht]
\renewcommand{\arraystretch}{1.5} 
\centering
\caption{Performance of Retrieval}
\label{tab:embedding_table}
\begin{tabular}{|c|c|c|c|}
\hline
\textbf{Embedding} & \textbf{Dataset} & \textbf{Hit Rate} & \textbf{MRR} \\
\hline
\multirow{4}{*}{bge-large-en-v1.5} & DSCC & 0.936 & 0.882 \\
& DUSD & 0.948 & 0.872 \\
& DUP  & 0.922 & 0.892 \\
& DUCC & 0.911 & 0.846 \\
\hline
\multirow{4}{*}{bge-base-en-v1.5} & DSCC & 0.896 & 0.869 \\
& DUSD & 0.920 & 0.861 \\
& DUP  & 0.923 & 0.890 \\
& DUCC & 0.927 & 0.875 \\
\hline
\multirow{4}{*}{bge-small-en-v1.5} & DSCC & 0.869 & 0.881 \\
& DUSD & 0.874 & 0.908 \\
& DUP  & 0.855 & 0.857 \\
& DUCC & 0.861 & 0.948 \\
\hline
\end{tabular}
\end{table}

\begin{figure}[htbp]
	\centering
	\includegraphics[width=2.8in]{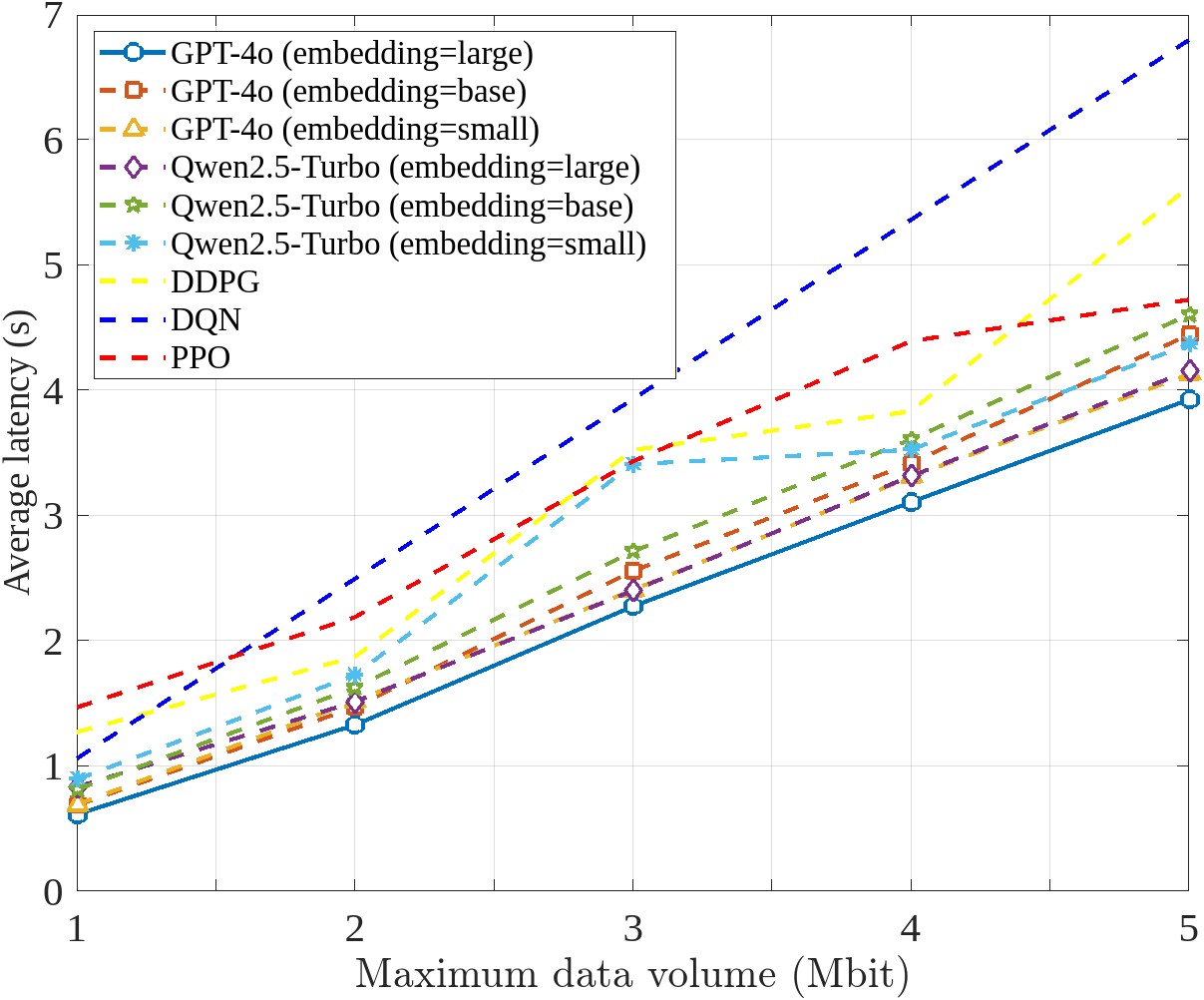}
	\caption{Performance comparison of different algorithms under the data volume $D_k$ of different computing tasks.}
	\label{fig:data_size}
\end{figure}

Fig. \ref{fig:data_size} shows the average latency versus the maximum data volume for various methods, including RAG approaches (GPT-4o and Qwen2.5-Turbo with different embedding sizes) and DRL-based methods (DQN, DDPG, PPO). It can be observed that the average latency increases consistently with the maximum data volume, which is due to the higher computational demands associated with larger datasets. Moreover, DQN, DDPG, and PPO methods have higher latency compared to all RAG-based methods. Among them, DQN demonstrates a significant increase in latency, as it relies on static optimization strategies and struggles to adapt to dynamic and complex user queries. Its single-threaded decision-making further exacerbates the computational bottleneck, leading to rapid latency growth as data volume increases. The RAG-based method uses a dynamic retrieval mechanism to retrieve the configuration information of the sending users based on their identity. This dynamic retrieval capability effectively reduces the processing time of irrelevant data and provides a reliable source of decision-making for task offloading.

\begin{figure}[htbp]
	\centering
	\includegraphics[width=2.8in]{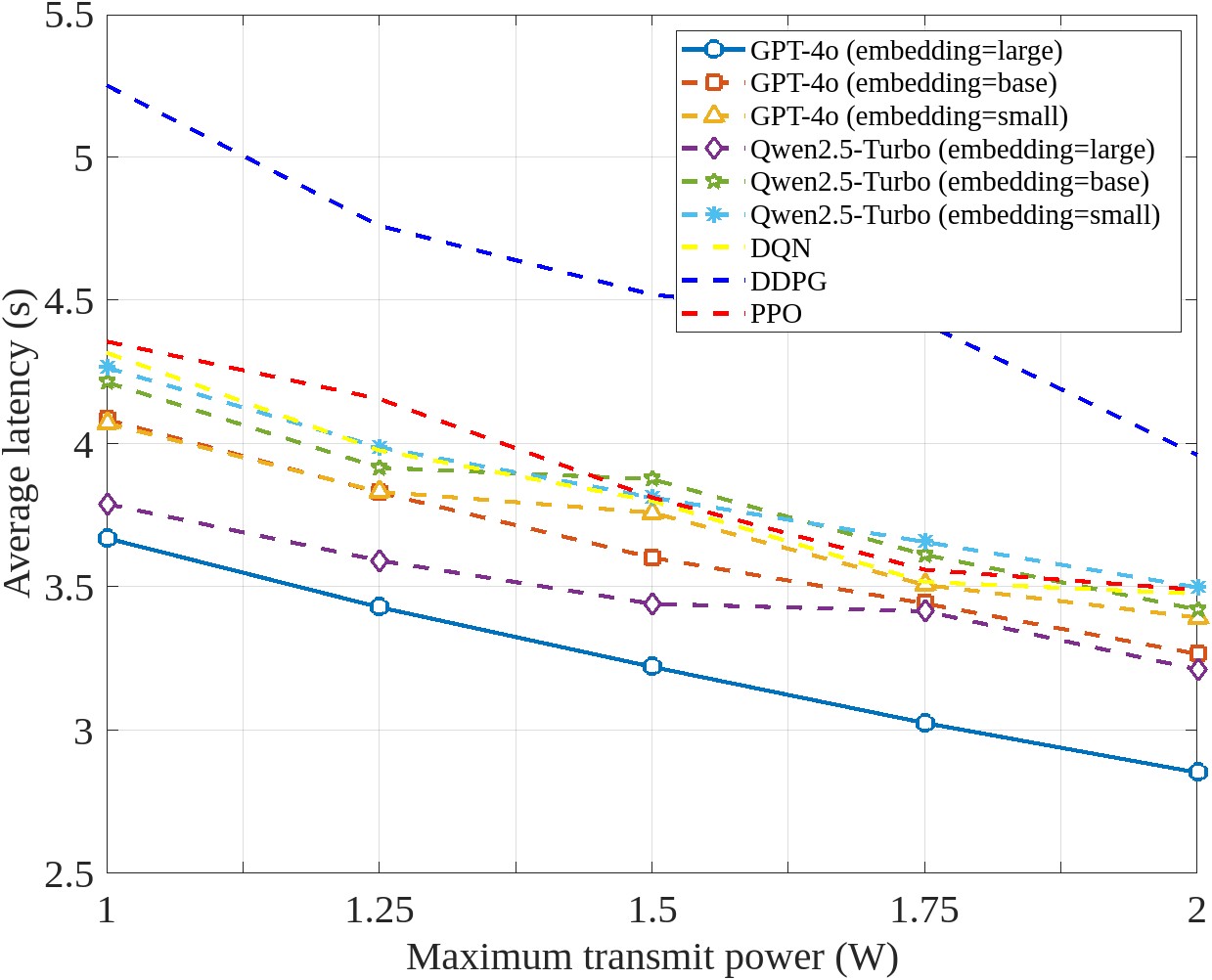}
	\caption{Performance comparison of different algorithms under the maximum transmit power $P_{max}$ of different MUs.}
	\label{fig:P_max}
\end{figure}

Fig. \ref{fig:P_max} shows the average latency versus the maximum transmit power for different methods. It can be observed that as the maximum transmit power increases, the average latency decreases across all methods.
In most cases, RAG-based methods outperform the baselines. The method of DQN directly benefits from higher transmission rates, as communication bottlenecks are reduced, and it does not require complex dynamic retrieval or parallel processing mechanisms to achieve noticeable improvements. This simplicity allows DQN to adapt effectively in environments where communication constraints are relaxed, leading to significant latency reduction. Similarly, PPO achieves comparable performance because of its robust policy optimization mechanism and ability to stabilize decision-making through balanced exploration and exploitation.  PPO's structure allows it to leverage increased transmit power more efficiently than DDPG, maintaining stable latency reductions.
On the other hand, DDPG performs the worst due to its reliance on continuous action space optimization and deterministic policy gradients, which fail to adapt to dynamic transmission environments.  Unlike DQN and PPO, DDPG struggles to exploit the benefits of increased power because its policy gradient optimization is more prone to instability in rapidly changing scenarios. Additionally, it cannot manage the discrete adjustments required for offloading tasks effectively.  Furthermore, DDPG's lack of dynamic retrieval and parallel decision-making mechanisms makes it inefficient in reducing latency when compared to the RAG-based approaches.


\begin{figure}[htbp]
	\centering
	\includegraphics[width=2.8in]{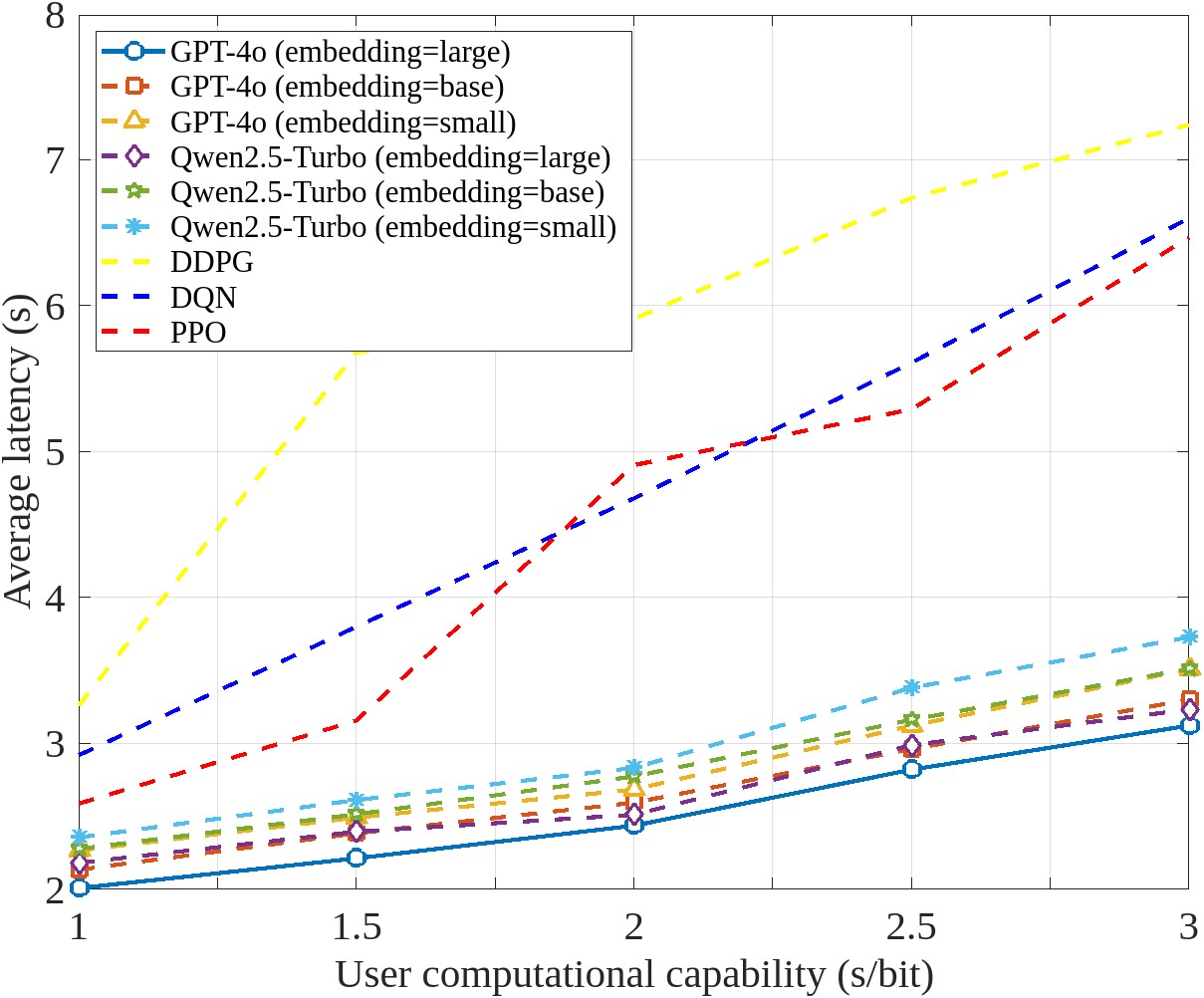}
	\caption{Performance comparison of different algorithms in different user computational capability $\frac{\phi}{f_k}$ s/bit.}
	\label{fig:user_f}
\end{figure}

Fig. \ref{fig:user_f} demonstrates the average latency versus the computing capability of MUs, which is measured in seconds per bit (s/bit). The horizontal axis represents the time required to compute one bit of data. As the computing capability (s/bit) decreases, the average latency consistently increases. The RAG methods, including various configurations of GPT-4o and Qwen2.5-Turbo, exhibit better performance, with their latency increasing at a slower rate compared to other baseline methods. In contrast, DDPG shows the worst performance, with its latency increasing as the computational capability decreases. 
This result is intuitive, as reduced computational capability results in slower data processing and increased reliance on the server. This shift in workload leads to bottlenecks in data processing and communication, causing higher latency. Furthermore, slower processing speeds amplify the impact of communication overhead, further contributing to the overall latency.
GPT-4o with large embeddings is able to adapt to scenarios with reduced computational capability, due to its advanced optimization mechanisms for workload management and communication. On the other hand, the sharp increase in DDPG, PPO and DQN's latency highlights its inefficiency in handling scenarios where both computation and communication resources are constrained. Configurations with smaller embeddings, such as Qwen2.5-Turbo, also show similar trends but are less efficient.

\begin{figure}[htbp]
	\centering
	\includegraphics[width=2.8in]{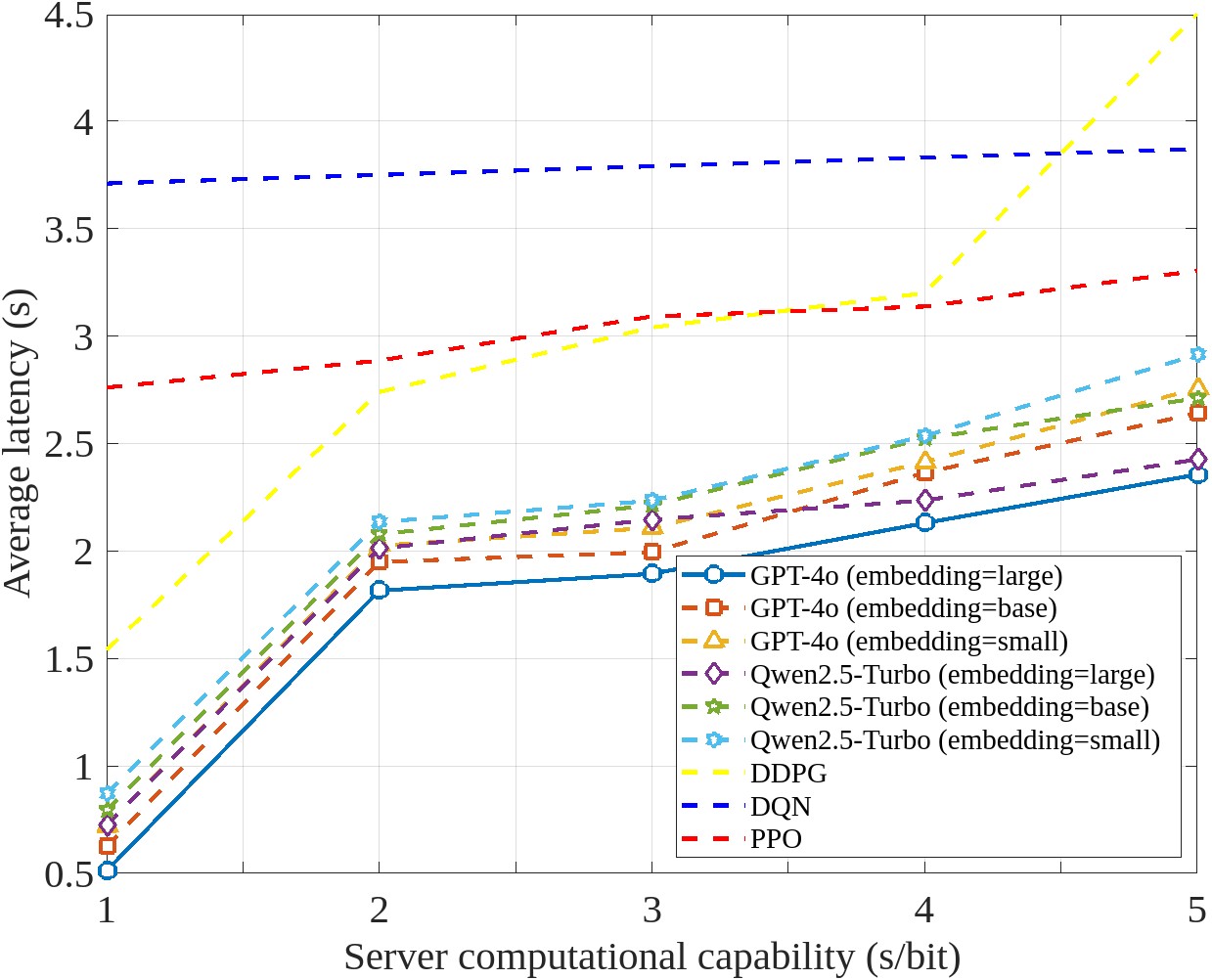}
	\caption{Performance comparison of different algorithms in different server computational capability $\frac{\phi}{F}$ s/bit.}
	\label{fig:server_f}
\end{figure}

Fig. \ref{fig:server_f} illustrates the average latency versus server computational capability, which is measured in seconds per bit (s/bit). The horizontal axis represents the computation time required to process one bit of data. As server computational capability decreases, the average latency increases. However, the trend differs depending on the computational capability range: when the computational capability is high ($1–2 \mathrm{e}^{-8}$ s/bit), the increase in latency is significant. When the computational capability is low ($3–5 \mathrm{e}^{-8}$ s/bit), the rate of increase in latency slows down.
The sharp increase in latency for high server computational capability ($1–2 \mathrm{e}^{-8}$ s/bit) is due to the server's inability to process tasks efficiently when resources are heavily constrained. At this stage, the server struggles to keep up with incoming tasks, leading to significant processing bottlenecks and queuing latency. As the computational capability decreases ($3–5 \mathrm{e}^{-8}$ s/bit), the server has sufficient capacity to handle the incoming tasks. Consequently, other system bottlenecks, such as communication overheads or transmission latency, begin to dominate the overall latency. Compared to the DQN, DDPG and PPO methods, the RAG-based methods consistently perform better, maintaining lower latency across all computational capability levels. Its ability to optimize resource usage and manage computational loads allows it to adapt well to varying server capabilities. 
In comparison, the performance of DQN, DDPG, and PPO is relatively poor. Specifically, at the beginning, DQN performs the worst. However, its latency increases steadily. As server computational capabilities gradually decrease, the performance of DQN surpasses that of DDPG. This may be because DQN operates on discrete value representations, which inherently reduce sensitivity to fluctuations in environmental parameters.

\section{Conclusion}

In this paper, we propose a novel RAG-based approach for computation offloading in MEC systems. Traditional DL methods have limitations in data dependency, generalization ability, and interpretability, making them difficult to adapt to the diverse requirements of MEC systems. By combining LLMs with a real-time information retrieval mechanism, our approach is able to obtain the latest system information and generate an adaptable offloading strategy that meets user-specific requirements. At the same time, RAG improves the interpretability and reliability of the decision-making process, providing a scalable solution for multi-user, multi-task, and highly dynamic offloading scenarios. Our study first applies RAG to computation offloading in MEC, providing a valuable reference for future research. Future work could explore the application of RAG in larger-scale MEC systems and strengthen user experience through LLMs.

\bibliographystyle{IEEEtran}
\bibliography{shortref}

\end{document}